\documentclass[lettersize,journal]{IEEEtran}
\usepackage{amsmath,amsfonts}
\usepackage{algorithmic}
\usepackage{array}
\usepackage[caption=false,font=normalsize,labelfont=sf,textfont=sf]{subfig}
\usepackage{textcomp}
\usepackage{stfloats}
\usepackage{url}
\usepackage{verbatim}
\usepackage{graphicx}
\usepackage{amsmath,amssymb}
\usepackage{braket}
\usepackage[utf8]{inputenc}
\usepackage[T1]{fontenc}
\usepackage{mathptmx}
\usepackage{etoolbox}
\usepackage{lineno}
\usepackage{chngcntr}
\usepackage{cite, float}
\UseRawInputEncoding
\usepackage{color, soul}
\def\BibTeX{{\rm B\kern-.05em{\sc i\kern-.025em b}\kern-.08em
    T\kern-.1667em\lower.7ex\hbox{E}\kern-.125emX}}
\usepackage{balance}
\begin{document}
\title{Compressive single-pixel read-out of single-photon quantum walks on a polymer photonic chip}
\author{Aveek Chandra, Shuin Jian Wu, Angelina Frank, James A. Grieve
\thanks{Manuscript received 23 May 2023; accepted 26 May 2023. Date of publication 1 June 2023; date of current version 16 June 2023. This work was supported in part by the Ministry of Education (MOE), Singapore, through AcRFTier3 under Grant MOE2018-T3-1-005, in part by the National Research Foundation, Singapore, and in part by A*STAR under CQT Bridging Grant.{(Corresponding author: Aveek Chandra)}}
\thanks{Aveek Chandra, Shuin Jian Wu, and Angelina Frank are with the Centre for Quantum Technologies, National University of Singapore, Singapore 117543
(e-mail: cqtavee@nus.edu.sg; cqtwsj@nus.edu.sg; angelina.frank@u.nus.edu).}
\thanks{James A. Grieve is with the Quantum Research Centre, Technology Innovation
Institute, Abu Dhabi 9639, UAE, and also with the Center for Quantum
Technologies, National University of Singapore, Singapore 117543 (e-mail:
James.Grieve@tii.ae).}
\thanks{Digital Object Identifier 10.1109/JPHOT.2023.3281830
}}

\markboth{IEEE PHOTONICS JOURNAL}%
{Compressive single-pixel read-out of single-photon quantum walks}
\maketitle 
\begin{abstract}

Quantum photonic devices operating in the single photon regime require the detection and characterization of quantum states of light. Chip-scale, waveguide-based devices are a key enabling technology for increasing the scale and complexity of such systems. Collecting single photons from multiple outputs at the end-face of such a chip is a core task that is frequently non-trivial, especially when output ports are densely spaced. We demonstrate a novel, inexpensive method to efficiently image and route individual output modes of a polymer photonic chip, where single photons undergo a quantum walk. The method makes use of single-pixel imaging (SPI) with a digital micromirror device (DMD). By implementing a series of masks on the DMD and collecting the reflected signal into single-photon detectors, the spatial distribution of the single photons can be reconstructed with high accuracy. We also demonstrate the feasibility of optimization strategies based on compressive sensing.

\end{abstract}

\section{Introduction}
\noindent
\IEEEPARstart{T}{HE} integration of quantum photonic circuits into chip-scale devices (QPICs)~\cite{Wang2020, Moody2021} is a key enabler of increasingly multifunctional and reconfigurable systems. In a photonic chip, light is confined in waveguides, propagating as guided optical modes. Engineering efforts over the last two decades have realized a diverse array of components such as on-chip couplers, interferometers, and resonators. In a quantum photonic device, these components are adopted and combined to implement gate-based devices and devices based on coherent quantum walks.

In quantum devices, it is frequently of interest to measure photon correlations at the output waveguide ports of the chip, to test the device and verify the desired outcome. This requires that light emitted from each output waveguide mode be collected and routed to single photon detectors, for example avalanche photodiodes (APDs) or superconducting nanowire single-photon detectors (SNSPDs)\cite{Crespi2013, Adcock2019}. While it is possible to couple each output to a discrete detector channel, this may be tedious, and quickly becomes unfeasible when the number of output ports is large, or the ports are densely spaced.\par

In this work, we demonstrate an inexpensive and versatile method to image and map the output modes of a multi-port chip by using a single pixel imaging (SPI) strategy~\cite{Duarte2008}. In the context of remote sensing, SPI with heralded single photons\cite{Johnson2022, Kim2021} has been shown to be noise-robust and loss-tolerant in detecting signal that would otherwise be obscured by strong background illumination. In our work, we investigate a heralded single-photon quantum walk on a polymer chip by single-pixel imaging of the intensity distribution at the output facet of the device.

Our method is low-cost and works both for imaging as well as measuring photon correlations among the output modes. The use of a click-based detector with high timing resolution (timing jitter $\sim$ 300 ps) enables us to distinguish pairs of photons from other events (e.g. detector dark counts, stray light, uncorrelated photons).
Additionally, we can extend our method to image a two-dimensional geometry of distributed optical modes in a multi-layer or three-dimensional architecture\cite{Lustig2022}. Finally, our approach can be simply adapted to image wavelengths spanning from visible to infrared, and a wide range of light intensities (from single photons to bright illumination), as long as an appropriate detector is operated below saturation.
While in principle multi-pixel devices such as EMCCDs (with a combination of short acquisition time and nanosecond triggering) could be deployed to similar effect~\cite{Alberti2017}, these devices are comparatively expensive and not well suited to identifying correlated events.

\par
Quantum walks (QWs) have become an integral tool for development of algorithms in quantum simulation\cite{Aspuru-Guzik2012,Portugal2013} and quantum computation\cite{Childs2009} as well as study of high-dimensional and reconfigurable graphs, disorder and topological phenomena in photonics\cite{Kitagawa2010, Kitagawa2012}. Recently, topologically protected modes have been demonstrated using waveguide arrays on a polymer chip\cite{Frank2022}. Our method could be of use in the implementation of quantum Boson sampling\cite{Spring2013} and quantum neural networks\cite{Zhao2021}, as well as opening avenues to explore complex topological geometries\cite{Poulios2014, Kondakci2017, Tang2018} in multi-dimensional, multi-port waveguide systems.
\section{Experimental Methods}
Light propagation in a waveguide array with nearest-neighbor coupling can be understood as continuous-time, discrete-space quantum walk on a 1D graph, with its Hamiltonian given by
\begin{equation}
\hat{H} =
 c\sum_{i=1}^{\infty}\beta_i a_i^\dagger a_i + \kappa_{i,i-1}a_{i-1}^\dagger a_i + \kappa_{i,i+1}a_{i+1}^\dagger a_i,
\end{equation}
where $\beta_i$ is the propagation constant, $a_i^\dagger$ and $a_i$ are the Bosonic creation and annihilation operators for photons in waveguide $i$, $\kappa_{i,i+1}=\kappa_{i,i-1}=\gamma$ are coupling coefficients that describe coupling between adjacent waveguides and $\gamma$ and $\beta_i$ are constant for uniform array.
For a single particle in waveguide $i$, $a_i^\dagger\ket0 =\ket1$ at initial time $t_0=0$ hopping to an adjacent waveguide $j$, $a_j^\dagger\ket0 =\ket1$ at time $t$, its transition amplitude is given by unitary evolution $U_{i,j}(t) =\bra1_i \exp(-i\hat{H}t)\ket1_j$. The single-particle transition probability, $p_{i,j}(t)=|U_{i,j}(t)|^2$ describes the photon intensity distribution across the waveguide array for a single-waveguide input~\cite{Meinecke2013}. Interestingly, this light intensity distribution can be shown to be the same irrespective of whether a single-photon state, a Fock state $\ket n$  or a coherent state is injected at the single-waveguide input\cite{Bromberg2009}.

At the heart of our method is a DMD, comprising a rectangular grid of micromirrors (called `pixels'), with the ability to orient each pixel to an `ON' or `OFF' state. In our case (model DLP LightCrafter$^{\text{TM}}$ from Texas Instruments), 415872 pixels are arranged in a 608 (columns) $\times$ 684 (rows) diamond pattern with each pixel having a diameter of 7.64 $\mu$m and a pitch (spacing) of 10.8 $\mu$m.
\noindent
\begin{figure*}
\centering
\includegraphics[width=11cm]{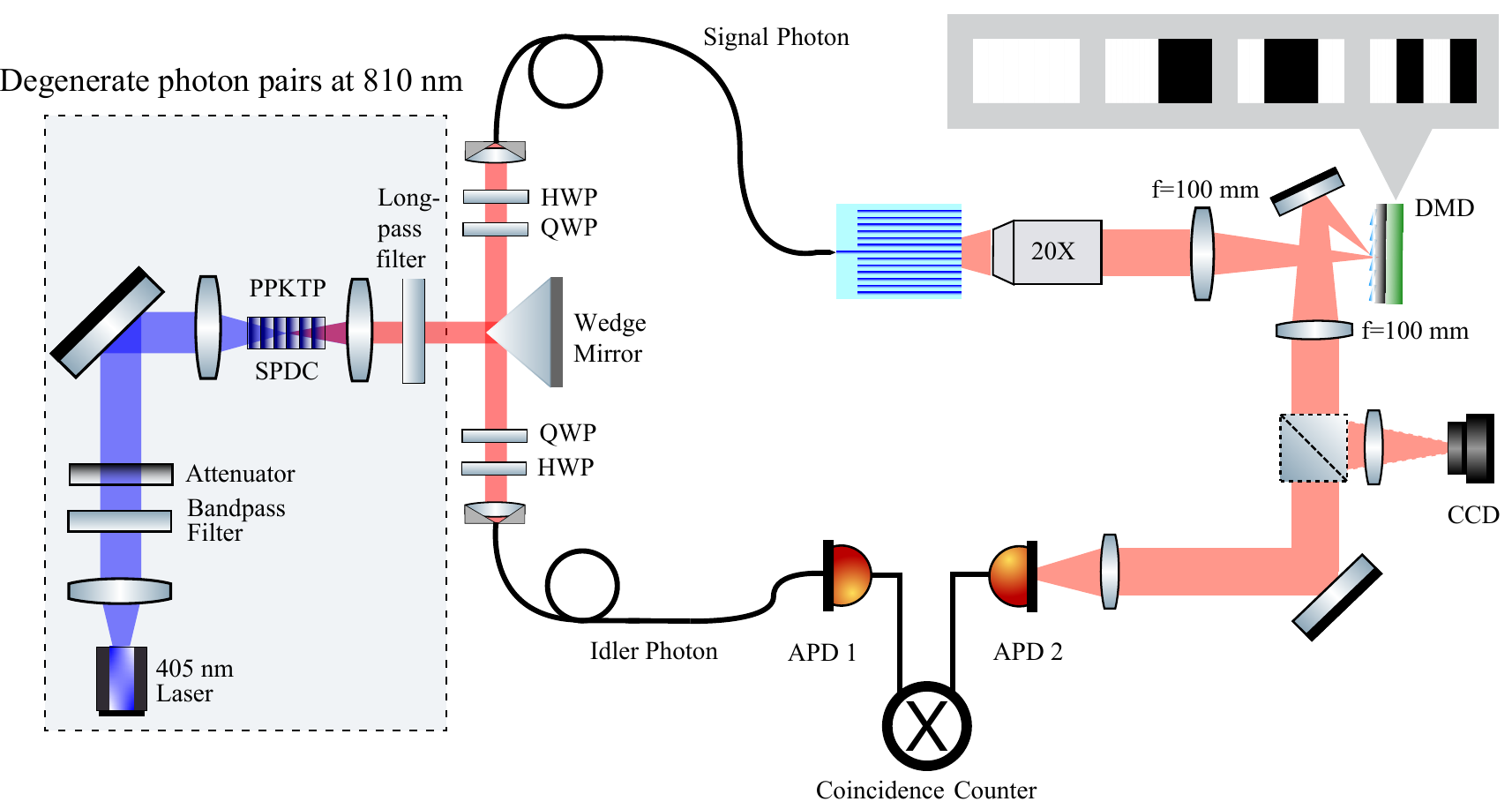}
\caption{Experimental setup. The degenerate photon pair source at 810 nm is shown by dashed box on the left. The photon pairs are split by a wedge mirror and coupled to respective SM fibers. The idler photon is sent to APD 1 while the signal photon is injected into the waveguide array chip via a lensed fiber, undergoing a quantum walk. The light from the chip endface is collected by an objective and focused on the DMD. The light reflected off the DMD is directed and focused on free space APD 2. For each projected pattern on the DMD, coincidences from both APDs are recorded. When laser light is injected for alignment purpose, the intensity distribution is viewed by a CCD with a beam-splitter inserted in the optical path shown by a dashed square.}\label{fig:setup}
\end{figure*}

The experimental setup is shown in Fig.~\ref{fig:setup} and the details can be found in Appendix \ref{app:methods}. Photon pairs (frequency-degenerate with center wavelength at 810\,nm) are produced via a type-0 spontaneous parametric down-conversion (SPDC) process by using a 10\,mm-long periodically poled potassium titanyl phosphate(ppKTP) crystal, pumped by a CW laser at 405 nm. The pair is split by a wedge mirror and are coupled into SM fibers, as described in \cite{Perumangatt2020}. One photon (`idler') from the pair makes its way to the fiber-coupled APD 1 while the other one (`signal') is injected into a polymer waveguide array via a lensed fiber. The waveguide device is fabricated in-house on a polymer platform as described in Appendix \ref{app:methods} and in ref.\cite{Grieve2017a, Frank2022a}. The polarization of both photons can be adjusted using waveplates positioned between the wedge mirror and fiber couplers.
Light emerging from array is collected by an infinity-corrected microscope objective (magnification $10\text{X}$), and imaged onto the active area of the DMD. 
Light reflected from the `ON' pixels of the DMD is directed and focused by lenses onto a free-space APD 2, which has an active area of diameter 180\,$\mu$m. Detection events from the APDs are encoded as electronic pulses, and routed to an AND gate, allowing us to accumulate ``coincident detection events'' in a counter.\par

The photon pair source produces about 2 million pairs/s per mW of pump power under optimal phase-matching conditions, at crystal temperature $25.7^{\circ}$C. The pairs/singles ratio (``heralding efficiency'') is maintained at 0.25. The photon pair source is at first characterized by sending both idler and signal photons to fiber-coupled APDs. Once it has been optimally calibrated, the fiber-coupled APD in the signal arm is disconnected and the signal photon is redirected to the waveguide array chip via a lensed fiber. 

In the experiment, 250\,$\mu W$ of pump power is used to produce $5\times10^5$ photon pairs per second at the source. Owing to 2$\%$ overall transmission, single-channel rates of $S1=1.9$ million events per second on the herald (idler) arm and rates of $S2=4500$ events per second on the signal arm are observed. The coincident event channel measures 1000 coincidences per second between the signal and the idler arms. With a coincidence window of approximately 16\,ns (constrained by the width of the digitized APD signal), the rate of accidental or background coincidences can be estimated to be $\sim$137 per second, resulting in a signal-to-noise ratio above 6. All event rates reported in this letter, unless stated otherwise, are uncorrected, with detection losses or inefficiencies included by default.\par
The overall system transmission i.e. the probability of a photon inserted into the chip finally impinging onto the detector is $\sim$2$\%$, representing 17\,dB loss. This figure is dominated by  two main sources: diffractive losses from the DMD (approximately -6\,dB), and insertion loss from the fiber to the waveguide (approximately -10\,dB). While our current experiment is constrained by these losses, there exist clear optimizations which could largely mitigate both, for example the adoption of tapered edge-couplers in our waveguide devices.\par

In the experiment, we are concerned with imaging 13 waveguide modes at the end face of our photonic chip. Light from these optical modes falls on a small section of DMD, a region corresponding to $64\times16=1024$ pixels. For our system magnification, each mode can be addressed by a few ($\sim$4) DMD pixels. We refer to these groupings as ``superpixels'', which may be switched between ON and OFF states to effectively mask- or gate-out individual waveguide modes.
Before this strategy can be deployed, it is necessary to gain precise knowledge on the location of waveguide modes in DMD pixel coordinates. This necessitates the reconstruction of the optical intensity profile at the DMD, which we perform using imaging techniques borrowed from the single pixel camera literature~\cite{Edgar2019, Gibson2020a}.

In single pixel imaging (SPI), spatial information about the image is reconstructed using a single ``bucket'' photo-detector and a large number of ``mask'' elements placed in a conjugate image plane between the detector and the image. The strategy is particularly useful where a specialized sensor is needed, and a large array of such sensors is impractical. As measurements must be taken in ``series'', acquisition times may be long in comparison to multi-pixel approaches, and scales linearly with the active area that must be imaged.

In our experiment, an image area corresponding to $64\times16=1024$ pixels have to be reconstructed using 1024 masks in total, with 2048 measurements required for full reconstruction (we must measure the signal from both positive and negative masks). At the single-photon level, acquisition time is constrained by signal-to-noise considerations -- for our devices 1 second gives acceptable performance. In total, the acquisition time for a full reconstruction is approximately 40 minutes.

To expedite this process and reduce the required acquisition time we implement compressive single-pixel imaging (CS-SPI), where only first $M$ out of $N$ rows of an optimally ordered set of masks (for example the well-known Hadamard basis) are used for image reconstructions. Though the acquisition here is not ``fast'' by imaging standards, here the image plane is relatively static (affected only by alignment drift). In general, we found it quite feasible to capture data over several hours, and this could be extended further by engineering a more stable fiber-to-chip coupling. 

While CS-SPI has been explored in the literature for natural images, this is to our knowledge the first demonstration of CS-SPI in the context of read-out of a photonic device.
In this work we investigate two orderings: the Cake-cutting~\cite{Yu2019b} and Russian Dolls~\cite{Sun2017} schemes, with the Cake-cutting strategy detailed below (the Russian Dolls ordering can be considered to be the same, with additional steps detailed in Appendix~\ref{app:ordering}).
\begin{enumerate}
\item Rows of Hadamard matrix, H$_{1024}$ are reshaped into $64\times16$ matrices in order to facilitate use as binary masks.

\item The Cake-cutting algorithm identifies and counts connected regions in a mask (i.e. a continuous block of either 1 or -1).

\item Masks are assigned a rank based on the number of constituent blocks and sorted in ascending order. See Appendix Fig.~\ref{fig:ordering} for an illustration.
\end{enumerate}

Intuitively, the Cake-cutting ``rank'' value can be thought of as approximately proportional to the spatial frequencies associated with each mask \cite{Vaz2020}. Consequently this represents an ordering in which the lower frequencies are sampled first, on the assumption that they contribute significantly to the signal compared to higher frequencies.

\begin{figure}[!b]
\centering
\includegraphics[width=9cm]{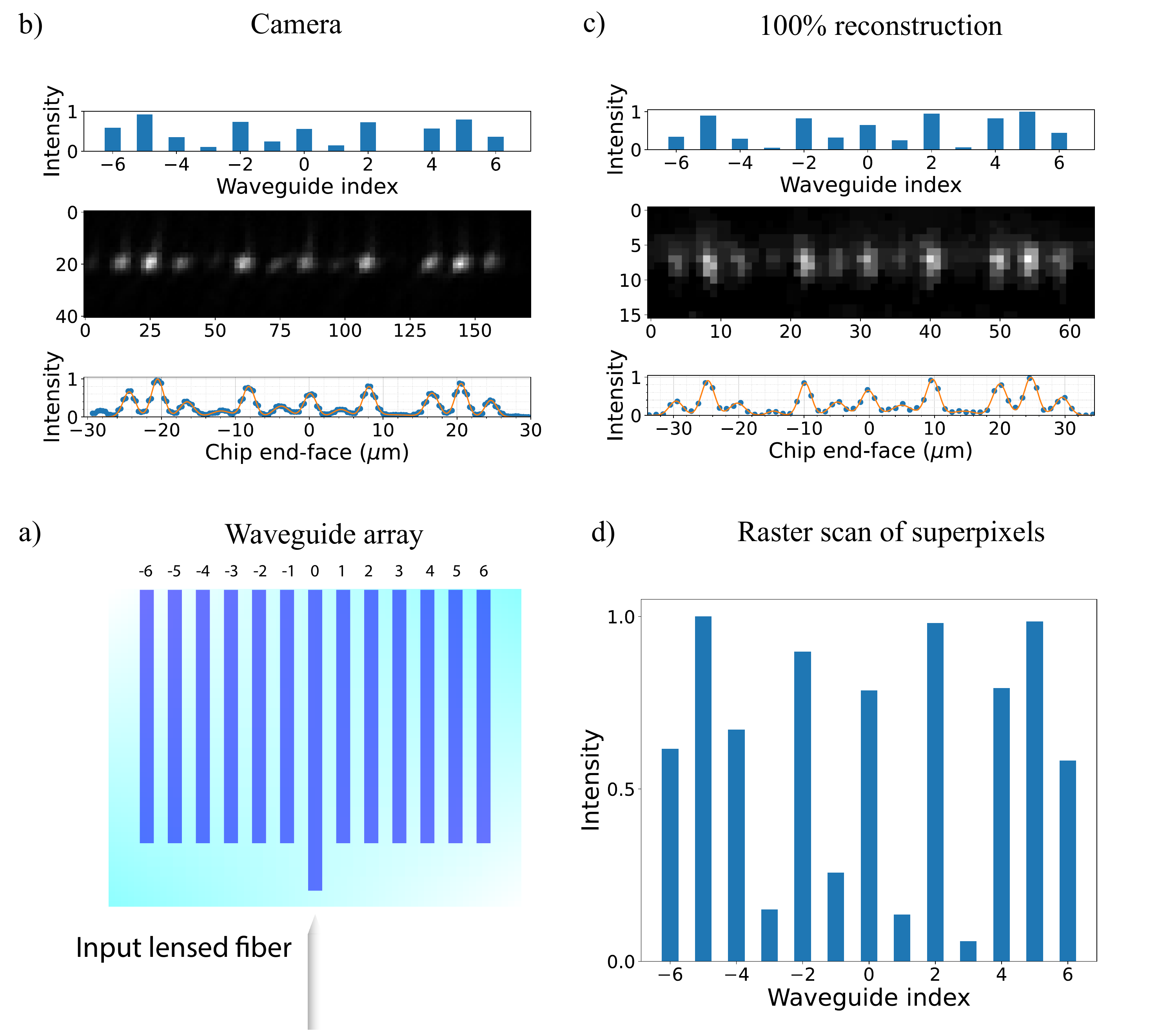}
\caption{(a) Schematic of waveguide array chip where light is injected into the middle of array at waveguide index 0 via a lensed tapered fiber. (b) CCD image and plots showing 13 visible modes, when laser light at 802 nm is injected into the array. (c) Full reconstructed image and plots of heralded single-photon quantum walk by SPI. (d) Reconstruction of heralded single-photon quantum walk by raster scan of superpixels.}\label{fig:combo}
\end{figure}
\begin{figure*}[!t]
\centering
\includegraphics[width=18cm]{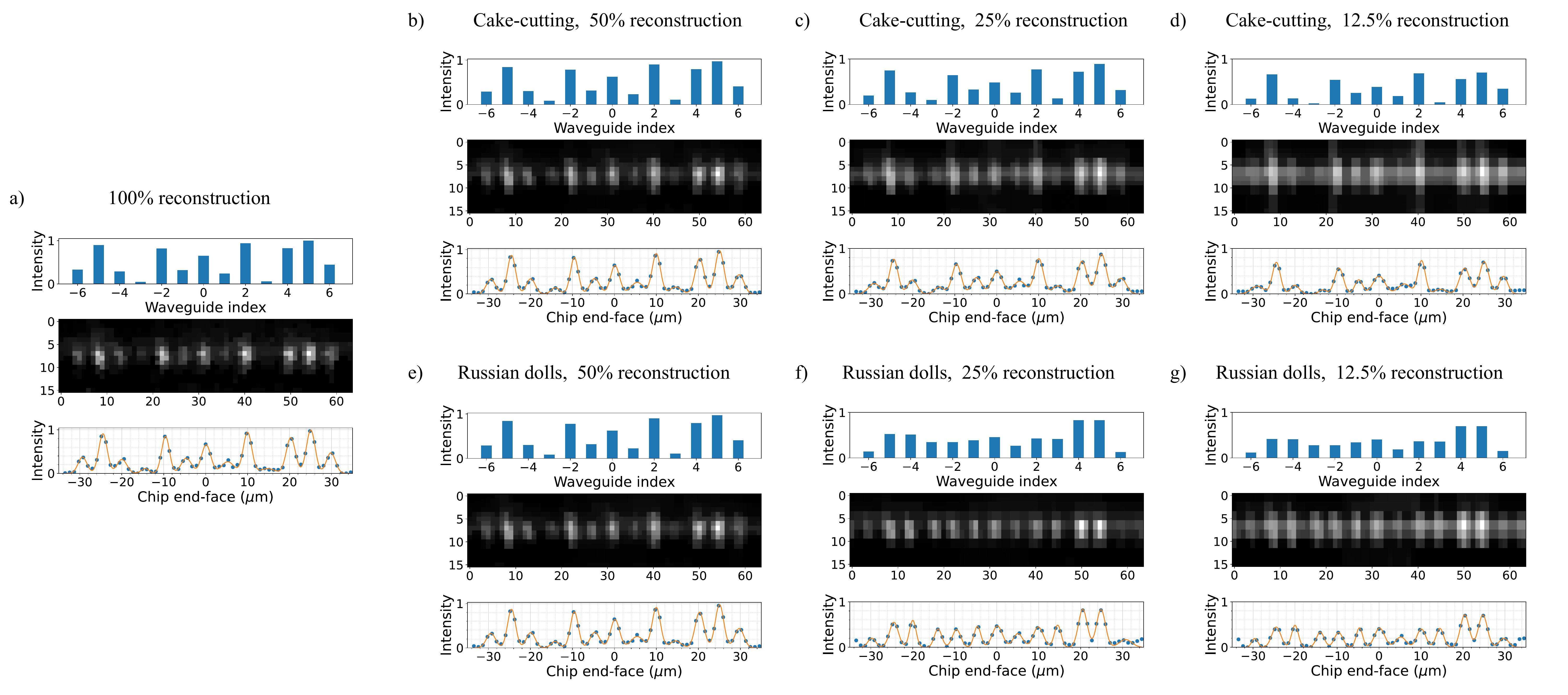}
\caption{Compressed Sensing Single Pixel Imaging (CS-SPI) of heralded single-photon quantum walks. This data uses the Cake-cutting ordering at reconstruction ratios (a) 100\%, (b) 50$\%$, (c) 25$\%$ and (d) 12.5$\%$.}\label{fig:compress}
\end{figure*}

We perform SPI of our waveguide chips as follows: for each mask implemented on the DMD, single and coincident event rates are recorded with 1\,s integration time. The order in which the masks appear is determined a priori by the algorithm mentioned above. The image is reconstructed progressively in real-time (using data collected so far) and recorded as a function of different reconstruction (compression) ratios (i.e. fraction of masks measured).
We make use of the total variation minimization by augmented Lagrangian and alternating direction algorithms, abbreviated as  'TVAL3'~\cite{Li2013}, an open source MATLAB package~\cite{Li2010} from Rice University that has been developed for CS-SPI. The algorithm is commonly utilized in the SPI community for its speed of execution and is known to obtain excellent image reconstruction with low mean squared error (MSE).

\section{Results and Discussion}
\noindent Single pixel imaging of the mode structure of a uniform waveguide array is summarized in Fig.~\ref{fig:combo}. We use a lensed fiber to inject light from an 802\,nm laser source into the central waveguide (index 0), as shown in Fig.~\ref{fig:combo}(a). The waveguides support both TE and TM spatial modes at zero order, and we ensure operation in the TE regime throughout this work. The nearest-neighbour coupling coefficient $\gamma$ is estimated to be
0.0085 mm$^{-1}$ for our array, with 3\,$\mu$m waveguide separation chosen to enable us to resolve individual modes at the output facet.
With bright light used as an input, the distribution of optical mdoes at the output can be directly visualized using a standard silicon CCD as shown in Fig.~\ref{fig:combo}(b). Constructive and destructive interference, characteristic of the quantum walk of a coherent-state input is evident in the symmetric distribution.

For a heralded single-photon input, the full reconstruction using single-pixel imaging is shown in Fig.~\ref{fig:combo}(c). The reconstructed images of modes are visible in a rectangular grid of 64 $\times$ 16 pixels. Also shown is the extracted intensity profile based on summation of intensities by column for the corresponding image. By fitting this data to a weighted sum of Gaussian functions, we extract a set of intensities $I$ for all modes, giving us the mode spectrum.
{Eq.~\ref{eqn:fid}} defines the Mean Squared Error (MSE) metric that we use to quantify similarity between two mode spectra represented by $I_{\text{orig}}$ (`original' dataset) and $I_{\text{recon}}$ (`recontructed' dataset):

\begin{equation}\label{eqn:fid}
\text{MSE} = \frac{1}{N}\sum_{i=1}^N (I_{\text{recon}} - I_{\text{orig}})^2,
\end{equation}
where $N$ refers to the number of intensity values compared (i.e. the number of modes), 13 for this chip. An MSE of zero would imply identical intensity profiles. The MSE can be used to compare reconstructed spectra with data captured using a multi-pixel device, assuming identical resolution and normalization. As the single-photon quantum walk is same as the quantum walk of a coherent state, we compare the fully reconstructed single-photon QW shown in {Fig.~\ref{fig:combo}(b)} with the camera-captured QW of a coherent state {Fig.~\ref{fig:combo}(c)}. The computed MSE is 0.02, and we believe some of this discrepancy is related to spectral characteristics of the different light sources used.
\begin{figure}[!b]
\centering
\includegraphics[width=9cm]{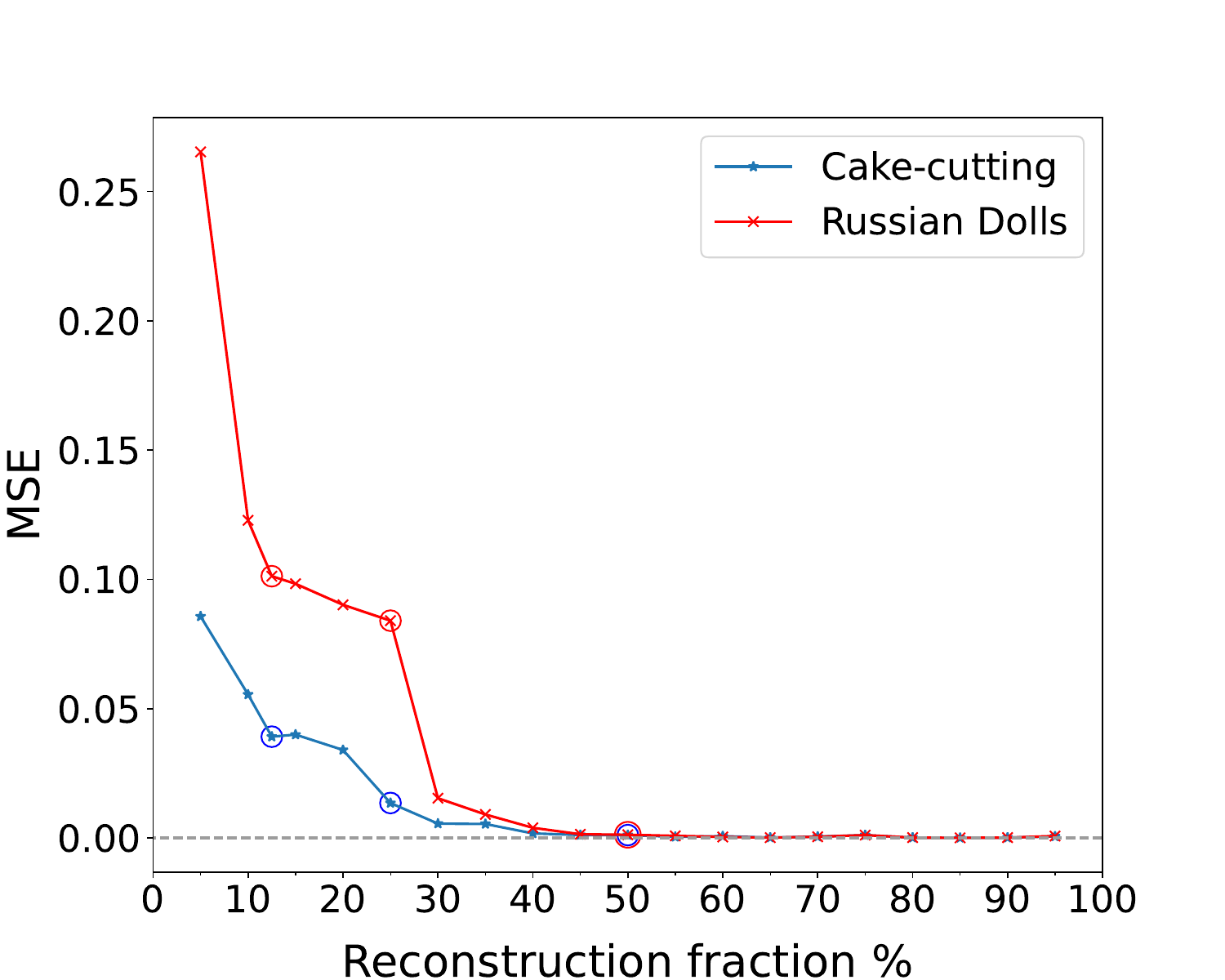}
\caption{MSE value computed at various reconstruction (compression) ratios for both Cake-cutting order and Russian Dolls order with points (b), (c), (d) from Fig~\ref{fig:compress} circled.}\label{fig:compress_fid}
\end{figure}
Fig.~\ref{fig:combo}(d) is the reconstruction of the same heralded single-photon quantum walk via raster scan of superpixels, in other words, switching 'ON' superpixels one-by-one to gate out individual waveguide modes. The data has been recorded the same way, except now the integration time of APDs is set to $10$ s for a enhanced signal to noise ratio. The datasets from Fig.~\ref{fig:combo}(c) and Fig.~\ref{fig:combo}(d) when compared, seem to agree fairly well with a computed MSE of 0.02. 
Though `superpixel raster scan' has been used here to verify the result obtained from single-pixel imaging method, it is clear that a reasonably high resolution reconstruction is a pre-requisite for this approach, in order to properly define the superpixel locations.

Fig.~\ref{fig:compress} shows the results of compressive single-pixel imaging (CS-SPI) using the Cake-cutting and the Russian Dolls orderings at various reconstruction ratios - (a) 100\%, (b) 50$\%$, (c) 25$\%$ and (d) 12.5$\%$. The data derived from partial reconstructions is compared quantitatively via the MSE, contrasting the partial reconstruction with the spectrum obtained using the full dataset. MSE is plotted against reconstruction ratio (as a percentage) in $5\%$ increments (approximately 51 masks) in {Fig.~\ref{fig:compress_fid}}, for both sorting strategies. At lower reconstruction (compression) ratios, the Cake-cutting order clearly outperforms the Russian Dolls order, with the MSE of Cake-cutting case remaining below 0.03 even at 25\% reconstruction.
At higher reconstruction ratios (50$\%$ and above), the basis orderings perform very similarly, with high fidelity (approximately zero MSE). We believe this is mainly due to the relatively large area of the optical modes (which are imaged onto multiple DMD pixels) and the relatively sparse information content of the reconstructed images. The second half (latter $50\%$) of the patterns in the sorted Hadamard set are generally associated with ``high spatial frequencies'', with a corresponding high block number (see Fig.{~\ref{fig:ordering}}. This implies that their use would increase sensitivity to information in the image that has a correspondingly fine structure. Since our modes are relatively large, there is not much additional information to be gained by sampling these patterns.

We expect the compressive sensing strategy to be particularly compelling where there are large number of output ports filling a larger region on the DMD. A $50\%$ compression reduces the acquisition time by 20 minutes per device in our case, and this reduction in time would scale roughly linearly with the number of outputs on the chip. In principle, raw acquisition time
can be improved by reducing the image magnification factor (such that the modes fall on fewer DMD pixels). This would speed up data gathering, but it might also be expected to reduce the degree of compression possible (i.e. it might require more than half the masks to be sampled). It would also be expected to increase the sensitivity of the read-out to optical alignment, so an overall saving in experimental time cannot be assumed. Fortunately in our case, the fiber-chip mechanical system is sufficiently stable to permit approximately two hours of data acquisition in a fully automated process.

In addition to the reconstruction of single-photon quantum walks, we have experimented with coupling two-photon states into the waveguide array (see Appendix \ref{app:twophoton} for details). While system losses did not permit the observation of non-classical correlations, we believe that this is a promising area of future research. There are several areas in which the losses could be reduced, including adopting inverse-taper edge couplers on the polymer platform. Improved transmission will facilitate single-pixel imaging of two-photon quantum walk and measurement of two-photon correlations not only in a simple one-dimensional array like ours but in complex, two-dimensional geometries.\par

\section{Conclusion}
\noindent
In conclusion, we have demonstrated the effective use of single-pixel imaging techniques in the read-out of weak optical signals at the end-facet of a waveguide chip. We further employ compressive sensing strategies to reduce the acquisition time. Although system losses currently restrict the technique, we believe that the scheme can be applied to the study of non-classical two-photon states, particularly if low-noise detectors (such as superconducting nanowire devices \cite{Wolff2020, EsmaeilZadeh2021}) are available. While we have chosen to focus on a one-dimensional uniform waveguide array, this strategy is clearly applicable for arbitrary waveguide geometries, for example the two-dimensional output facets of direct-written chips \cite{Poulios2014, Meany2015, Tang2018}. For QPIC devices with a large number of output modes, the reduction in acquisition time afforded by the compressive sensing (CS-SPI) approach will be particularly important. Finally, our method allows for addressing individual output modes by definition of ``superpixels'', which allows the system to be dynamically reconfigured for single-output measurements. It is our hope that this powerful new application of single pixel imaging can open up new opportunities in chip-scale quantum devices.

\section*{Acknowledgement}

\noindent The authors would like to thank Alexander Ling and his team for all the support and fruitful discussions. Filip Auksztol, Paul Thrane and Adithyan Radhakrishnan contributed to the early stages of the project. We thank Mohamed Riadh Rebhi, Isa Ahmadalidokht and David Phillips for discussions too. All experimental work was carried out at the Centre for Quantum Technologies, National University of Singapore.

\appendix

\subsection{Methods}\label{app:methods}
\noindent \textbf{Photon pair source} - Our photon pair source is constructed using the geometry described in ref.~\cite{Perumangatt2020}. Degenerate photon pairs are produced via Type 0 spontaneous parametric downconversion (SPDC) in a 10\,mm periodically poled potassium titanyl phosphate crystal (PPKTP, Raicol Crystals Ltd.) with poling period 3.45\,$\mu m$ in a single pass configuration (see Fig.~\ref{fig:setup}, dashed box). 405\,nm light from a stabilized laser diode (ONDAX) is cleaned via a fluorescence bandpass filter and focused to a 120\,$\mu$m waist at the center of the PPKTP crystal. A long pass filter (at 736\,nm) and dichroic mirror placed after the crystal remove residual pump light. Photon pairs are produced degenerate around 810\,nm with typical 20\,nm bandwidth, with identical polarization. Pairs are separated into two paths using a wedge mirror, with each path coupled into a single-mode fiber. Quarter-wave and half-wave plates in each path enable independent polarization control.

To reconstruct a heralded single-photon quantum walk, one photon (`idler') is directly connected to a fiber-coupled APD, while the other (`signal') is coupled into the waveguide array by means of a lensed fiber (OZ Optics 630-HP). For the signal photon, polarization is set to be horizontal at the waveguide array. Before performing the single-photon experiment, light from a laser diode is coupled into the waveguide array to facilitate fiber-to-chip alignment. A beamsplitter in the optical path (shown as a dashed square in Fig.~\ref{fig:setup}) is inserted to image part of the light onto a CCD camera. With the CCD sensor and DMD located in conjugate focal planes, the position of the chip and microscope objective can be fine-tuned to achieve focusing of light on the DMD as shown in Fig.~\ref{fig:combo}(b). For single-pixel imaging (SPI), the beam splitter is removed from the path and all light is directed to either a PIN photodiode (for bright light) or APD (for single photons).


\noindent \textbf{Polymer waveguide} -
The waveguide chip is produced in-house following the processes described in~\cite{Grieve2017a, Frank2022a}. It is composed of two polymers: Gelest OE50 (Mitsubishi Chemicals) and Sylgard 184 (Dow) with refractive indices of 1.50 and 1.41 respectively.
Gelest OE50 (the high refractive index layer) is spin-coated to a thickness of 1\,$\mu$m onto a ``mold'' written into a photoresist via UV-lithography. A matrix of Sylgard 184 (low refractive index layer) with a thickness of $\sim$2 millimeters is then applied on top and cured at $70 ^\circ C$ for one hour. The structure is removed from the mold by immersion in dimethyl sulfoxide for several hours. The refractive index difference and waveguide dimensions are designed to achieve single-mode guiding.

When working with evanescently coupled arrays, light is edge-coupled into a ``feeder'' waveguide, which extends from the array to the chip edge as shown in Fig.~\ref{fig:combo}(a). The array comprises of uniform straight waveguides with 3\,$\mu$m pitch and length of approximately 9\,mm. Light in the waveguides propagates as a fundamental TE mode with coupling coefficient $\gamma$ estimated to be 0.0085\,mm$^{-1}$.
\begin{figure}[!t]
\centering
\includegraphics[width=8cm]{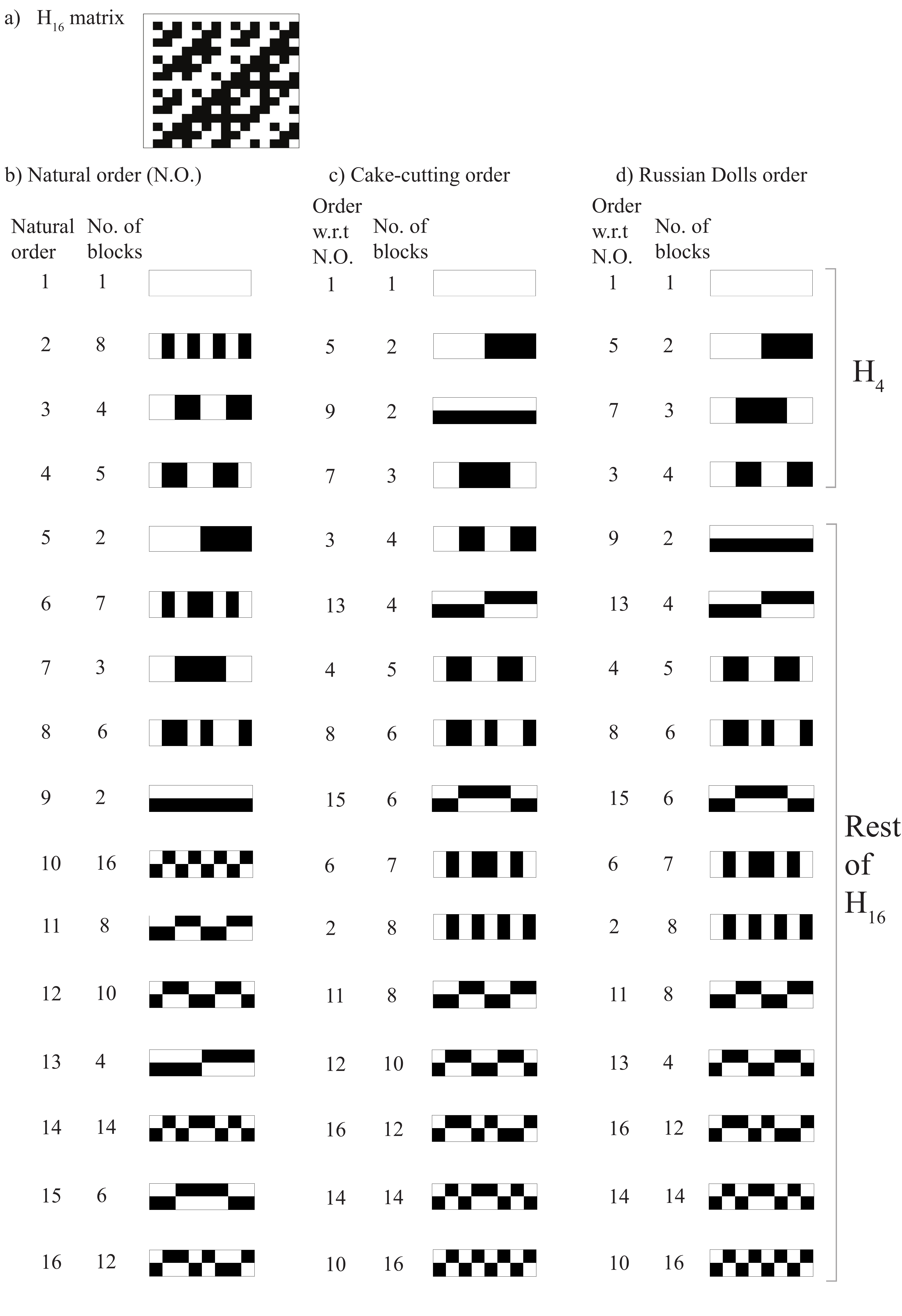}
\caption{(a) A $16\times 16$ Hadamard matrix, with elements 1 and -1 represented in white and black respectively. Each row of the matrix is transformed into a $8\times2$ mask in 2D and a complete set of 16 basis patterns, along with no. of blocks of connected regions, are presented for (a) Natural order, (b) Cake-cutting order and (c) Russian Dolls order.}\label{fig:ordering}
\end{figure}

\subsection{Ordering of Hadamard basis}\label{app:ordering}
In our experiment, for full reconstruction of the 16x64  pixel image  we employed 1024 masks in total. To illustrate how different basis orderings work, an example is helpful. Beginning with a $16\times 16$ Hadamard matrix, we choose the lowest order matrix of rank $8\times2$ -- see Fig.~\ref{fig:ordering}. Each row of the matrix is reshaped into a $8\times2$ array with the connected regions (or blocks) of 1 and -1 counted and shown alongside. Elements containing 1 and -1 are shown in white and black respectively for ease of visualization. The complete set of 16 basis masks along with the number of connected blocks are shown for the natural order in Fig.~\ref{fig:ordering}(b), the Cake-cutting order in Fig.~\ref{fig:ordering}(c) and the Russian Dolls order in Fig.~\ref{fig:ordering}(d). As mentioned previously, the Cake-cutting order is based on sorting the arrays by the number of blocks, in ascending order. 

Fig.~\ref{fig:ordering}(d) shows the Russian Dolls order, where the masks are split into H$_4$ and H$_{16}$ groups, with each group subsequently sorted based on number of blocks. Here, all Hadamard matrices, from lowest rank to the given rank are grouped in a library of basis patterns. For a matrix $H_{2^{2n}}$, there are lower order matrices $H_{2^{2n-1}}$, $H_{2^{2n-2}}$ and so on.
Each mask is compared with masks in this library and ones that are identified with that of the lowest ranked matrix will appear first, followed by masks that are identified with matrix of next higher order and so on. Masks not associated with any lower order matrix appear at the end. Finally, the masks in each of these ordered groups of matrices are sorted in ascending order.


\subsection{Two photon quantum walks}\label{app:twophoton}
\begin{figure}[!t]
\centering
\includegraphics[width=5cm]{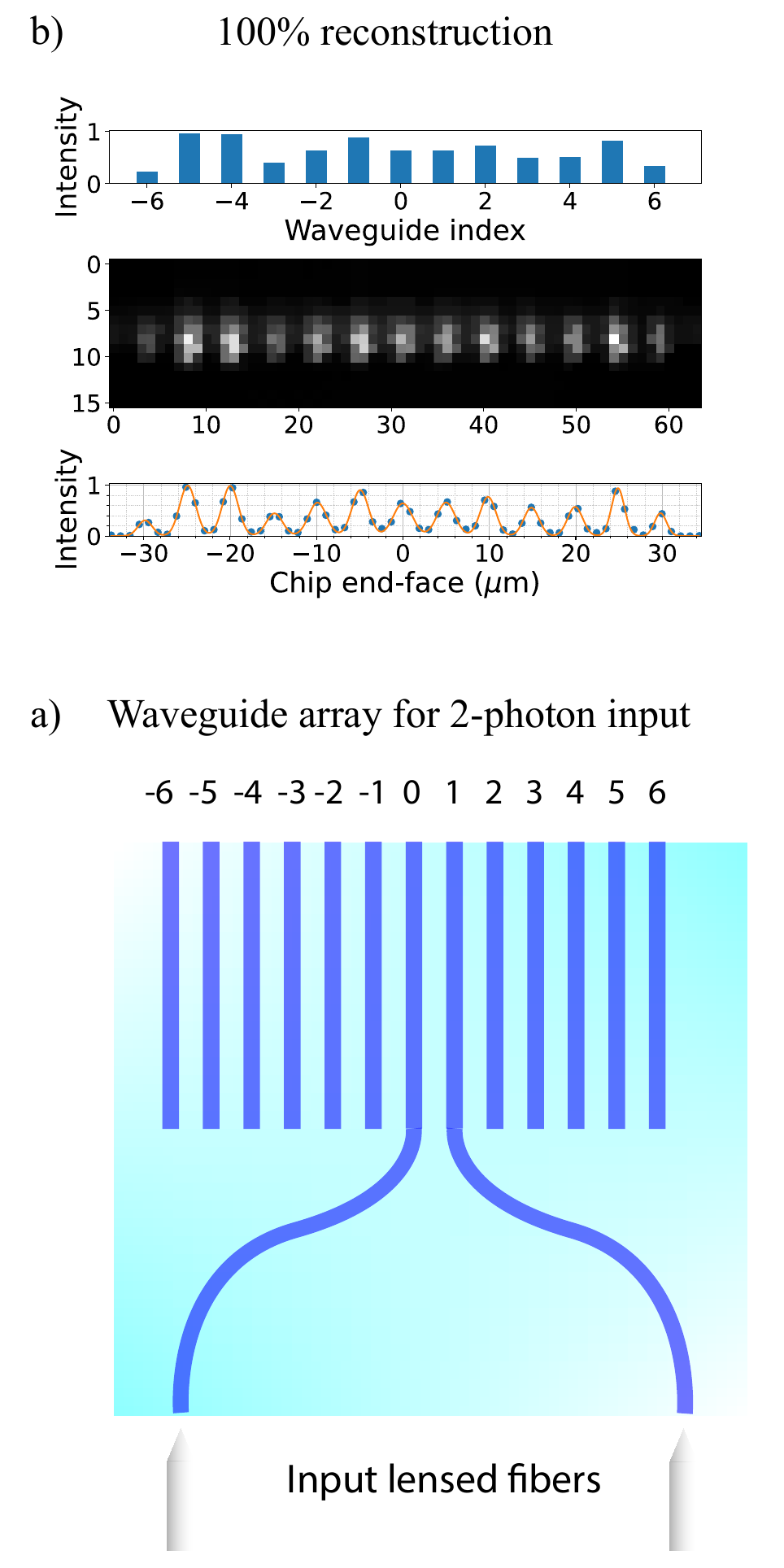}
\caption{a) Schematic of waveguide array chip for two-photon input where photons are injected into two adjacent waveguides, indexed 0 and 1, via lensed tapered fibers. (b) Full reconstructed image and plots of quantum walk for this two-photon input, the photons being degenerate in frequency and both polarized in horizontal direction. The observed intensity distribution is a convolution of two individual single-photon quantum walks and a two-photon quantum walk.}\label{fig:2ph}
\end{figure}
\noindent We explored the possibility of injecting two-photon states into adjacent waveguides via lensed fibers, as shown in Fig.~\ref{fig:2ph}(a). In order to address two waveguides independently by optical fiber, it was necessary to introduce a ``fan out'' region to the input side of the chip. Two central waveguides, indexed 0 and 1, are curved adiabatically over 8\,mm until they achieve a pitch of about 221 $\mu m$, that is compatible with a pair of optical fibers held by a v-groove.

Two lensed fibers are used to inject a pair of frequency-degenerate photons with horizontal polarization (i.e. the photons are indistinguishable). The output of the chip is again imaged using the SPI system, this time with the inclusion of a non-polarizing beamsplitter allowing us to route the masked output onto two independent avalanche photodiodes, after which the experiment proceeds as in the heralded case (Fig.~\ref{fig:setup}), with coincident detection events accumulated in electronic counters. Fig.~\ref{fig:2ph}(b) shows the reconstruction of modes from these two-photon events. The observed intensity distribution is described by the convolution of two individual single-photon quantum walks and a two-photon quantum walk.

Due to system losses (estimated at $\sim40$\,dB for the successful propagation of two photons), accidental coincidences (noise) are on the same magnitude as `true' two-photon events. A signal-to-noise ratio slightly above unity is neither sufficient to observe the two-photon quantum walk, nor does it allow for measuring non-classical two-photon correlations by pairwise selection of superpixels.
As discussed in the Experimental Methods, system losses could be reduced by using e.g. tapered edge-couplers on the waveguide chip. One can also redesign the post-DMD portion of the optical path to reduce ``field stop'' effects, however this was not feasible using our existing hardware. While it is clear that this method could be a powerful tool for investigating such two-photon systems, further refinements to reduce this loss will be required.

\bibliographystyle{IEEEtran}


\end{document}